# Non-equilibrium condensation of the first Solar System solids

Sebastien Charnoz´ [1], Jer´ ome Alˆ eon´ [2], Marc Chaussidon[1], Paolo A. Sossi[3], Yves Marrocchi[4], and Patrick Franco[1]

[1]Universite Paris Cit´ e, Institut de physique du globe de Paris, CNRS, 1 rue Jussieu, Paris, F-75005,´ France

[2]Institut de Mineralogie, de Physique des Mat´ eriaux et de Cosmochimie, UMR 7590, Sorbonne´ Universite, Museum National d'Histoire Naturelle, CNRS, Paris, 75005, France´

[3]Institute of Geochemistry and Petrology, ETH Zurich, Clausiusstrasse 25, Z¨ urich, 8092, Switzer-¨ land

[4]Universite de Lorraine, CRPG, CNRS, UMR 7358, 54000 Nancy, France´

[5]Corresponding author : charnoz@ipgp.fr

**Summary**

Primitive meteorites (chondrites) consist of an out-of-equilibrium assemblage of minerals formed during the assembling of our Solar Nebula [1]. The conditions under which their precursors condensed remain unclear as a result of subsequent re-processing in the protoplanetary disk or in asteroidal parent bodies. Chondrites are classified into three main classes enstatite (EC), ordinary (OC), and carbonaceous (CC) distinguished by different bulk composition and oxidation state [2]. While equilibrium condensation models explain the composition of some of their refractory components [3,4], they do not explain the emergence of three mineralogical classes. Moreover, the low pressures, steep temperature gradients, and short dynamical transport timescales in forming protoplanetary disks likely hindered equilibrium. Here we test the hypothesis that chondrite precursors formed via kinetic non-equilibrium condensation. Using a new time-dependent condensation model, we show that varying the cooling rate and pressure produce only three types of mineralogies. Departure from equilibrium yields increasingly oxidized and hydrous mineralogies. When projected into a Urey–Craig diagram, the predicted mineralogical types fall close to the redox states of EC, OC, and CC chondrites. These results suggest that the mineralogical diversity of chondrites may reflect, in part, local condensation kinetics, offering an alternative to large-scale variations of oxidation conditions.

## An history of condensation

The mineralogy of the oldest objects in primitive meteorites, the calcium-aluminium-rich inclusions (CAIs), appears to be successfully explained via an equilibrium condensation sequence (ECS) from a gas of solar composition [4,5]. These calculations are performed assuming that the gas and condensed phase(s) are able to equilibrate with each other at any given temperature and pressure. For temperatures in the range of 1300–2000 K the mineralogical sequence found in the ECS is similar to those observed in fluffy CAIs and some amoeboid-olivine aggregates (AOA) [3]. However, because CAI minerals (e.g., hibonite, melilite, spinel, grossite, Al–Ti-rich diopside, perovskite) vanish from the ECS below ∼1400 K [6], it has long been recognized [4,7,8] that some degree of disequilibrium is required to explain their coexistence with lower-temperature condensates (e.g., olivine, orthopyroxene, Fe–Ni alloy, forming at ≤1350 K at $10^{-4}$ bar). Early models invoked fractionated condensation, where solids formed in equilibrium were removed from the gas [9], leading to the introduction of *ad hoc* isolation factors [7,8,10] though what determines their precise value remains unclear [6].

In this context, the origin of the broad chemical variations between different classes of chondrites [2] remains uncertain. Enstatite chondrites possess iron entirely in their reduced (metallic) form, whereas increasing amounts of FeO are present in ordinary and particularly carbonaceous chondrites. While these variations may reflect, in part, secondary processes in their parent bodies, they also trace differences in the mineralogy of the initial condensate precursors [11]. However, ECS models, even when assuming non-solar gas compositions, fail to reproduce the observed range of mineralogical diversity[12] (Supplementary Sections 1 and 2). An alternative hypothesis states that each class of chondrites formed under different redox conditions [5,13]: silicates in carbonaceous chondrites would have condensed at oxygen fugacities ($fO_2$) near the iron–wustite (IW) buffer¨ [14], while the components of enstatite chondrites would have formed under highly reducing conditions. Yet generating such extreme local variations of $fO_2$ within the solar nebula remains a major problem. The protosolar gas is intrinsically highly reducing (H/O > 2000, H/C > 3000), making significant oxidation shifts physically unlikely. Mechanisms such as midplane enrichment in oxygen-rich dust [4,10,15] or inward transport of carbon-rich gas from the outer disk [14] have been proposed, but the required levels of enrichment are extremely difficult to achieve under realistic disk
conditions [13,15]. Because this paradox is based on the premise of equilibrium condensation, here we examine the effect of non-equilibrium condensation, as a natural consequence of low-pressure and vigorous dynamical environments, on the chemical properties of nebular condensates.

## Exploring non-equilibrium condensation

Non-equilibrium condensation is expected in protoplanetary disks. AOAs in carbonaceous chondrites show strong enrichments in light Si isotopes, consistent with rapid condensation over ~0.01 years under kinetic control [16]. Similarly, light tellurium isotopes in CC chondrules suggest fast, non-equilibrium condensation of chondrule precursors [17]. Because reaction rates decrease with temperature and pressure, equilibrium becomes increasingly unlikely at greater distances from the star—where temperatures are lower—or above the disk midplane, where pressure drops exponen-
tially [18].

The collapse of the molecular cloud onto the Solar Nebula involves complex dynamics [19–21], including heating near the star or in accretion shocks, followed by large-scale redistribution across a wide density range. In a minimum-mass solar nebula, pressures range from $\sim$0.1–1 bar inside 0.1 AU to $\sim 10^{-5}$–$10^{-7}$ bar beyond a few AU (Supplementary Section 4). At 1 AU and $P \sim 10^{-4}$ bar, 10$\mu$m grains grow in weeks to months [22], comparable to local free-fall times. Beyond 1 AU, collision rates fall [23], and equilibration times exceed disk lifetimes—or even the age of the Universe (Supplementary Figure 12). These constraints motivate study of non-equilibrium kinetic condensation.

For this purpose, we have developed a kinetic condensation code KineCond (detailed in Methods section and Supplementary Section 3). It computes the time-dependent condensation and evaporation of a gas of solar composition during cooling from 2000 to 130 K at constant pressure $P$ and during cooling timescale $T_c$ in a closed system. Gas–gas reactions are maintained in equilibrium, whereas gas–grain reactions proceed kineticaly. The reaction network couples 39 congruant condensation/evaporation reactions (Supplementary Section 3.2, Supplementary Table 2) and 38 gas–mineral nebular reactions (Supplementary Section 3.4 and Table 3) using an operator-splitting scheme with adaptive time stepping that conserves mass ( Supplementary Figure 5). Minerals grow by direct condensation balanced by evaporation, while mineral nebular exchange enables mineral transformations. We explore pressures $P$ from $10^{-9}$ to $10^{-2}$ bar and cooling timescales $T_c$ from 0.01 to 1000 years. Due to lack of experimental data we varied nebular reaction rates (activation energies) to simulate fast (FNR), moderate (MNR), and slow reactions rates (SNR), see Methods and Supplementary Section 3.4). Evaporation efficiencies ($\gamma$) are set to 0.1, in the range of experimental constraints [24]. To compare different condensation conditions, we define an empirical parameter $X = \log_{10}(T_c/\mathrm{year}) + \log_{10}(P/\mathrm{bar})$. The minerals formed in representative kinetic condensation sequences (KCS) for $-6 \leq X \leq 0$ appear in Figure 1 ( additional cases from $-11$ to $+1$ appear in Supplementary Section 6.)

Equilibrium vs. kinetic regimes

The KCS approximates ECS behavior under long cooling timescales, high pressure, and fast nebular reactions (FNR). This is illustrated in Figure 1.a (P = $10^{-3}$ bar, $T_c$ = 1000 years X=0). From

2000 K to ∼800 K, the KCS closely resembles the ECS (see ECS reference in Supplementary Figure 1), with Ca and Al rich minerals condensing first, followed by metallic iron and enstatite forming preferentially over forsterite. In contrast to ECS, below 800 K KCS mineralogy remains largely unchanged due to the exponential decline in reaction rates (Arrhenian kinetics), effectively isolating the condensates from the gas and preserving intact minerals typical of CAIs, even at low temperature. This outcome is achieved naturally in KineCond without invoking an *ad hoc* isolation factor [7,8,22]. Water ice condenses at ∼180 K.

While ECS and KCS show broad agreement for X>–5 and T>800K, key differences persist. The condensation sequence of high-temperature CAI-relevant minerals varies with nebular reaction rates (FNR vs. SNR; Figures 1.a,d), though corundum remains the first condensate. Plagioclase (anorthite, albite) forms in ECS but not in KCS: in ECS, it condenses below ∼1400 K (or 1200 K at $10^{-3}$bar), whereas in KCS, Ca and Al are already locked into higher temperature minerals by the time temperatures allow plagioclase formation. This reflects a kinetic barrier absent in ECS, where all elements are assumed to remain available regardless of temperature. In contrast, KCS captures the progressive chemical depletion of the gas under non-equilibrium conditions.

## Formation of three mineralogical types

The general agreement between KCS and ECS at high pressure and long cooling time breaks down when condensation occurs at low pressure or for short cooling time intervals (X<-5). Figures 1.c and 1.f presents the case P=$10^{-6}$ bar and $T_c$=1 year (i.e X=-6). After the condensation of hightemperature minerals (down to $\sim$ 1200K), and major silicates (1000K <T< 1200K), a rich variety of minerals condense for T<1000K. Iron is present in a variety of oxidation states in coexisting phases, including iron metal (Fe), fayalite ($Fe_2SiO_4$) and magnetite ($Fe_3O_4$). Phyllosilicates also condense (here, greenalite and lizardite). When cooling is fast enough ($X \leq -5$) anorthite, melilite (here akermanite and gehlenite) and spinel, three important components found in CAIs form (Fig 1.b to e).

We have led a systematic exploration of the KCS for different $P$ and $T_c$ (displayed in the forms of mosaics in Supplementary Section 6). Contrary to intuition, KCS does not evolve smoothly with $T_c$ and $P$ (i.e., $X$), but instead exhibits sharp transitions, defining only three distinct mineralogical types: Type A (Figure 1.a, 1.d), Type B (Figure 1.b, 1.e), and Type C (Figure 1.c, 1.f). Type A appears for $X > -5$ (upper-right triangle in matrix-mosaics displayed in Supplementary Section 6), corresponding to high-pressure, slow-cooling conditions and a mineralogy akin to enstatite chondrites (metallic iron and enstatite-rich). At the other extreme, Type C occurs for $X < -5$ (lower-left triangle in matrix-mosaics displayed in Supplementary Section 6), corresponding to low-pressure, fast-cooling, and characterized by oxidized iron phases (fayalite, magnetite, troilite, phyllosilicates) coexisting with high-temperature CAI minerals (hibonite, grossite) and various silicates. Type B ($X = -5$) is transitional (along diagonals in matrix-mosaic displayed in Supplementary Section 6) with Fe condensing as metal, troilite, and fayalite from ∼1150 K, at higher temperature than in ECS (around 650–500 K). Notably, each type shows limited internal variation despite large changes in $P$ and cooling time $T_c$.

Varying the activation energies of nebular reactions by several orders of magnitude has little effect on the resulting mineralogy (compare left and right columns of Figure 1 and matrix mosaics displayed in Supplementary Section 6). This suggests that phase formation is primarily controlled by the condensation of supersaturated species, with gas–grain reactions playing a secondary role despite the lack of experimental kinetic data (Supplementary Section 3.4.2).

Oxidizing material in fast cooling processes

At low pressure and short cooling timescales, mineral diversity results from the temperaturedependent availability of atoms in the gas, governed by condensation kinetics. Disequilibrium arises when cooling outpaces condensation, causing the gas to temporarily retain its high-temperature composition despite the low temperature. In the extreme case where the gas remains solar at any temperature (as a proxy for ultra-fast cooling), instantaneous condensation and evaporation fluxes are shown in Figure 2 (corresponding supersaturation coefficients are provided in Supplementary Section 7). At $\sim$1600 K, Ca and Al atoms can condense into CAI-like minerals (corundum, hibonite), while Fe, Mg, and Si condense into olivine and metallic iron near $\sim$1300 K at $10^{-5}$ bar. Full condensation of iron gas atoms takes 0.1–1 year (for 10 $\mu$m grains). If cooling is faster, some iron atoms remaining in the gas can later condense as fayalite or troilite near 1150 K and 1110K. In extreme disequilibrium, where significant Fe, Si, and Mg atoms still remain in the gas below 1050 K, oxidized phases such as magnetite and phyllosilicates (greenalite, lizardite here) condense directly from the gas. This contrasts with earlier models [9] in which Fe is entirely locked in metal, preventing the formation of later Fe-bearing phases. In our simulations, the amount of residual gas-phase Fe is dictated by condensation kinetics at high temperature (Figure 2), allowing greater mineral complexity as disequilibrium increases (Figure 1), particularly for Types B and C ($X \leq -5$). Finally, water ice only forms in Types A and B whereas in Type C, oxygen is stored in O–H bonds within phyllosilicates rather than as crystalline $H_2O$ ice (Figure 1.a,b,d,e).

## Condensation paths in the Urey-Craig diagram

We use the three chondrite classes—EC, OC, and CC—as a conceptual framework for exploring redox evolution and bulk mineralogy. While this simplification overlooks the full diversity of the sixteen recognized chondrite groups (such as the distinction between H, L and LL chondrites), it aims to elucidate differences in precursor material rather than reproduce the exact properties of each subgroup. The evolution of iron oxidation in our condensates is shown in the Urey–Craig diagram (solid lines in Figure 3a,b). All condensation sequences start in the lower-left corner and end on the solar Fe/Si line once condensation reaches 100% (reflecting our assumption of closed solar-composition system). Despite wide variations in pressure and cooling times ($-6 < X < 0$), final condensates cluster into three regions along this line, qualitatively matching the redox states of EC, OC, and CC chondrites (Figures 3.a,b). Type A mineralogies contain no oxidized iron and are dominated by enstatite over olivine and sulfides (<1 wt%), qualitatively consistent with EH chondrites. Type B paths starts near the EL field and progressively incorporate Fe into fayalite. The fast-reaction cases end near H chondrites when condensation is complete, while slowreactions case retain more metallic iron. Type C trajectories are more complex: they pass near L–LL fields, then through CO–CV regions, and end very oxidized, with Fe mostly in fayalite, magnetite, troilite, and phyllosilicates (greenalite). While their final states are less oxidized than classical CM or CI values, they are very close to the revised CI estimate (A25 in Figure 3) from [25]. Higher oxidation states (>0.7) are not reproduced by KineCond and likely result from parentbody alteration [26]. Because chondrite groups also exhibit non-solar bulk compositions (which origin is not invesigated here), we next explore how varying gas composition affects condensation paths using non-solar mixtures reflecting Fe/Si, Mg/Si, and Al/Si measured in different chondritic sub-groups (Supplementary Table 1 in Supplementary Section 9). Varying gas composition modifies the final oxidation state of the condensates (Figure 3.c), but all trajectories remain clustered within the Urey–Craig diagram. Slow condensation always produces reduced Type A assemblages, whereas rapid cooling yields oxidized Type C assemblages—independent of bulk gas composition. Condensation of L and LL-like gas compositions at $X = -5$ end closer to the observed L and LL fields but still remain systematically too reduced (see orange and grey star markers in Figure 3.c). Thus, our model captures at first order only the three chondrite classes (EC,OC,CC) in the Urey– Craig diagram, but not their subgroups, may be due to missing physics (i.e. silicate/metal sorting due to aerodynamic drag, C/O variations at the disk scale, open-system, gas–solid separation, or parent-body alteration), indicating that kinetic condensation alone cannot explain the full diversity of chondrites.

## Apparent $fO_2$ of kinetic condensates

The Urey-Craig diagram, by itself, does not reflect the oxygen fugacity ($fO_2$) of the environment in which the chondrites formed, because, the components of primitive chondrites are unequilibrated.

Estimates for the reequilibrated $fO_2$ of chondrites have been made using different techniques [27,28]. Enstatite chondrites record nominal $fO_2$s consistent with that of a gas of solar composition (about ΔIW ∼-6 to -7, where IW is the Iron-Wustite buffer). CI chondrites are more oxidized (with IW in¨ the range -4 to 0) than expected during condensation of a solar gas, unless temperatures fall below 400 K [27]. The leading hypothesis for their oxidation is the circulation of aqueous fluids on their parent bodies [26].

To estimate the apparent $fO_2$ of the non-equilibrium mineral assemblages A, B and C, the corresponding equilibrium $fO_2$ was calculated at T = 1500K and P=$10^{-4}$ bar using the *Factsage* software (Supplementary Section 8 and Supplementary Figures 19 and 20). Types A, B, C show re-equilibrated bulk compositions with distinct $fO_2$ trajectories as a function of temperature. When all condensable material is in solid form, types A, B and C show re-equilibrated $fO_2$ of ΔIW ∼4, ΔIW ∼-2, and ΔIW∼-0.5, respectively, spanning the range of estimated $fO_2$ in the different classes of (equilibrated) chondrites [14,28]. These equivalent $fO_2$s deviate from the true $fO_2$s at which these minerals condensed (the gas remains at ΔIW∼-6 to -7 over most of the temperature range; Supplementary Figure 20). Type C mineralogies are especially oxidized, and also have a water mass fraction of about 1-2 wt. %, without appealing to addition (i.e. super-solar) of oxygen or water to the system. High-temperature minerals, condensed at T> 1200K in KineCond, still have -6 < ΔIW < -10, at low temperature, close to the true $fO_2$ of their formation environment.

From precursors to the chondrite components

While KineCond redox trends broadly align with the three chondrite classes, whether kinetic condensation also shaped specific components, particularly chondrules, remains uncertain. These underwent significant reprocessing: melting, recycling, gas–grain interactions, and aqueous alteration—prior to their incorporation into meteorites [29]. Nevertheless, some chondrule mineralogies remain qualitatively consistent with a kinetic condensation origin.

Ferro-magnesian chondrules, the most abundant type, are classified based on their olivine-topyroxene ratio (> 80% olivine for type A, 20–80% for type AB, and <20% for type B) and Fe oxidation state (FeO < 5 wt.% for type I, FeO > 5 wt.% for type II) [30]. Type I chondrules typically contain abundant Fe metal and are thought to form under reducing conditions. In KineCond, slowly cooled Type A condensates ($X \simeq 0$) yield forsterite-rich mineralogies with Fe metal—similar to AOAs and other Type IA precursors [31,32].

Faster-cooling Type B condensates yield more oxidized assemblages with fayalitic olivine and residual metal, broadly resembling Type II precursors. This suggests that oxidation could result from disequilibrium condensation rather than the addition of external oxidants (e.g., CI-like dust or ice) [12]. Although KineCond does not resolve the full chemical complexity of chondrules, it suggests a coherent framework where redox diversity arises from local variations in pressure and cooling rate. The model also predicts coexistence of Fe metal, FeS, and oxidized phases (Figure 1, Supplementary Figure 6), consistent with mineral assemblages in chondrules [23,33]. Simulations of transient heating events, such as in bow shocks [34], followed by recondensation always converge to one of the three end-member mineralogies, depending on the cooling rate (Supplementary Section 10). Matrix materials likely reflect both condensation and parent-body alteration. However, primary amorphous silicates—thought to be initial matrix phases—cannot currently be modeled due to missing thermodynamic data [26], and their mineralogy has been extensively modified by thermal and aqueous processes [35].

Astrophysical context

Were Solar Nebula conditions suitable for producing Type A, B, and C mineralogies? This depends on how interstellar gas accreted onto the protoplanetary disk, a process still debated. Figure 4 outlines three proposed scenarios. In 1D collapse models conserving angular momentum [36–39], gas accretes near the proto-Sun where pressure and temperature are highest (see SI 4), favoring Type A mineralogies close to the star and Type C farther out. However, slow viscous spreading [38,39] limits cooling rates, likely preventing significant non-equilibrium effects. More recent 2D and 3D simulations [19–21,40] reveal infall extending out to $\sim$10 AU in asymmetric patterns. Accretion shocks above the midplane reach $T > 2000$ K [20,21], allowing thermal processing before gas cools and condenses into Type A, B, or C mineralogies (Figure 4.b) in the midplane. Additionally, outflows near the proto-star may eject hot material that cools rapidly in low pressure environment, promoting non-equilibrium condensation (Figure 4.c).

We propose that mineral precursors condensed under the conditions described here, and that their initial mineralogy influenced the outcome of subsequent disk and asteroidal evolution processes.

Observations of Class I protoplanetary disks reveal radial gaps that can act as dust barriers and limit radial mixing [41–43]. In our Solar System, the early spatial and temporal separation of NC and CC reservoirs supports this idea [44,45]. Such separation may have helped preserve the distinct mineralogies and bulk compositions of the precursors that gave rise to the three chondrite classes.

While based on simplifying assumptions—due to computational limits and scarce experimental data (see Methods)—our model suggests that non-equilibrium condensation may have contributed to the redox diversity of precursors and the first-order division of chondrites. These kinetic effects likely operated alongside other processes not included here, such as C/O variations, gas–solid separation, physical sorting, and parent-body alteration. Given that chondrite classes also differ in bulk composition, which our closed-system approach does not capture, our framework should be viewed as a conceptual basis rather than a full explanation.

Figures

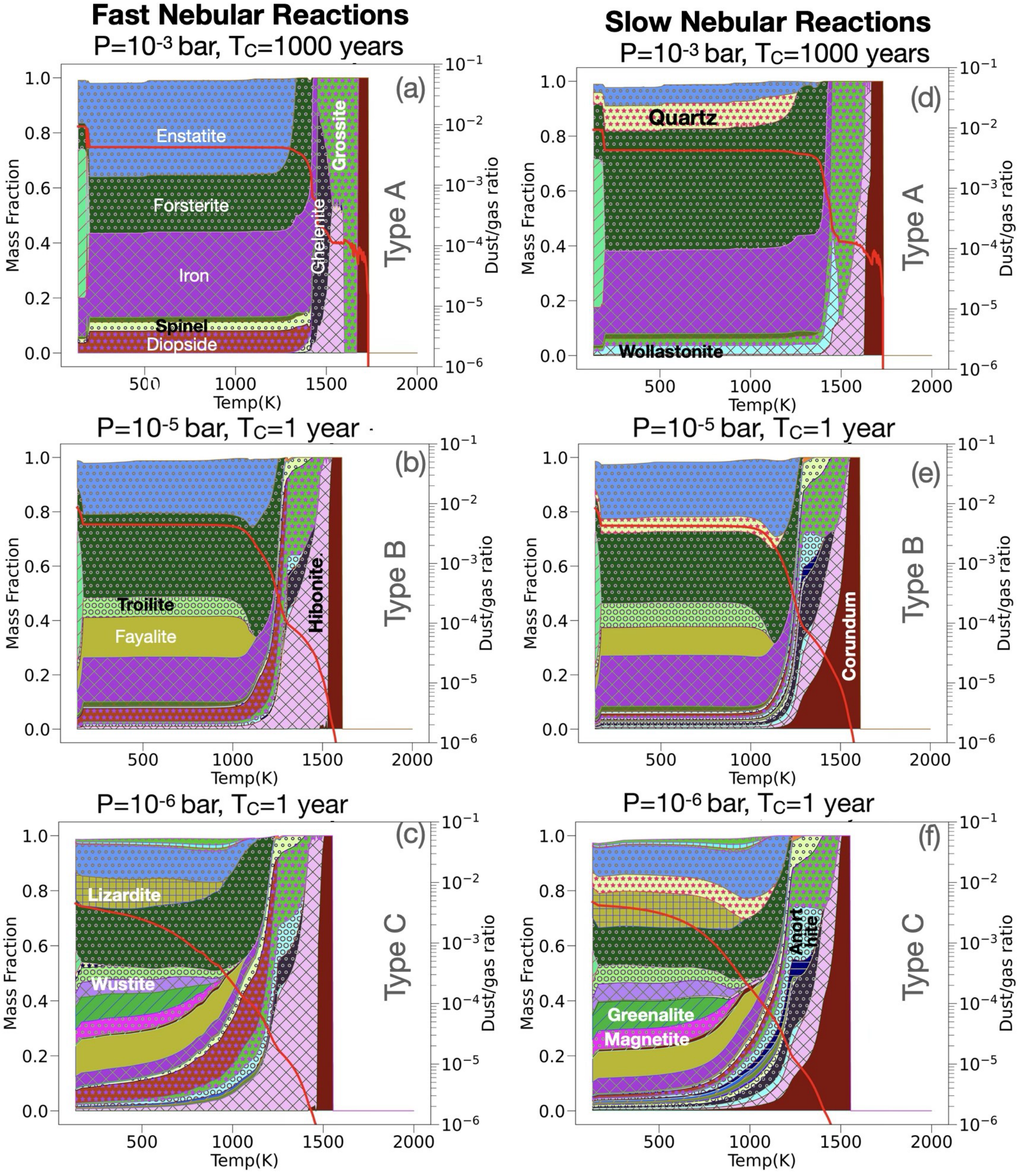

Figure 1: Example of kinetic condensation sequences. Resulting mineralogical sequences calculated for fast (FNR, left column) and slow (SNR, right column) nebular reaction models. Each panel shows the evolution at fixed pressure $P$ and cooling time $T_c$ (values above panels), corresponding to $X = 0$ (top: a,d), $X = -5$ (middle: b,e), and $X = -6$ (bottom: c,f). Colored regions indicate the solid mass fraction of each mineral (left axis) versus temperature; only phases exceeding 1% of total solid mass are shown. The red line denotes the dust/gas mass ratio (right axis). See Supplementary Figure 17 for color legend. Mineralogical types A, B, and C are discussed in the text. For $T < 700$ K, olivine/(olivine+pyroxene) mass ratios are 32%, 65%, and 66% for FNR cases (a–c), and 81%, 63%, and 66% for SNR cases (d–f).

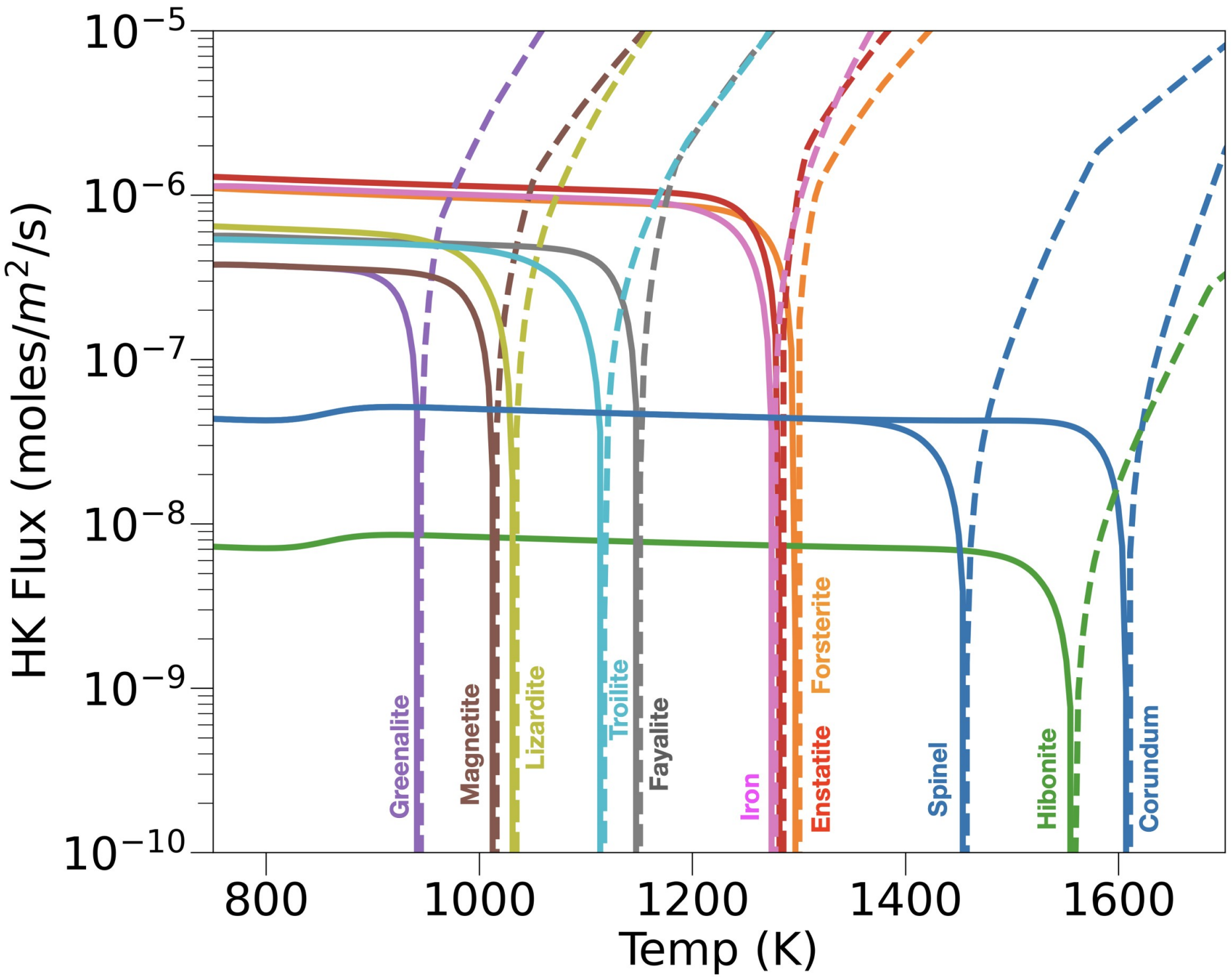

Figure 2: Hertz-Knudsen (HK) flux for a selection of minerals . Fluxes are displayed as a function of temperature, in a gas maintaining its solar composition (P=$10^{-5}$ bar). Solid line: condensation flux; dashed line: evaporation flux. A mineral condenses when the evaporation flux drops below the condensation flux, corresponding to super-saturation S>1[46] .

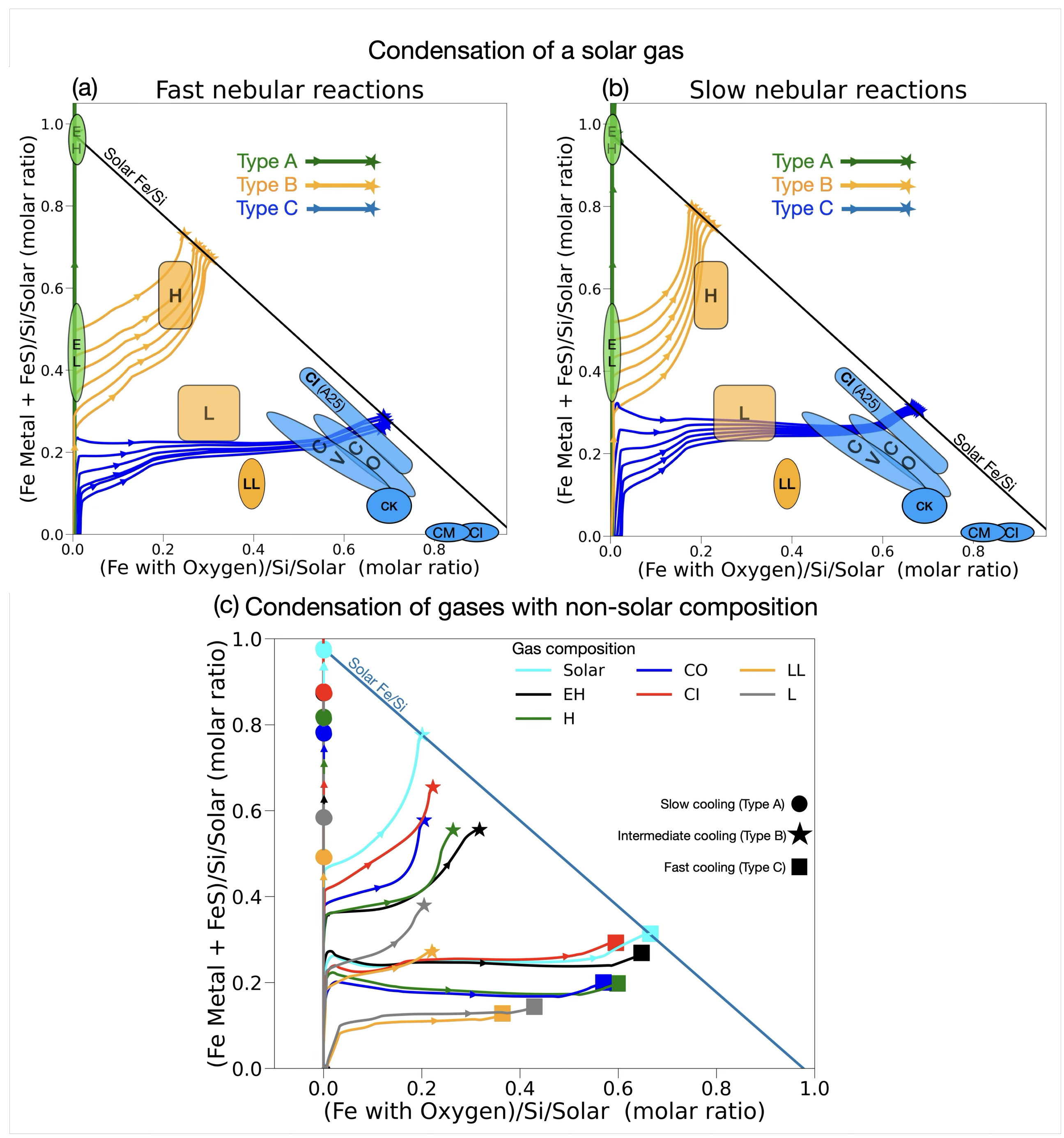

Figure 3: Evolution of condensates in the Urey-Craig diagram. This diagram shows the iron redox state of condensates along the condensation path calculated with KineCond, compared with chondrite compositions (colored shapes). The x-axis shows molar fraction of oxidized Fe, and the y-axis shows Fe in metal and sulfide, normalized to Si and to the solar Fe/Si ratio. Colored solid lines trace the redox evolution during condensation process of Types A, B, and C for various ($P, T_c$) cooling conditions. (a) Condensation paths of a solar-composition gas with fast nebular reactions (FNR). (b) Same as (a), but for slow reactions (SNR). All trajectories end on the solar Fe/Si line. (c) Condensation paths for non-solar gas compositions, using moderately fast reactions.

Colors correspond to different gas compositions and marker shapes denote different cooling rates ($X = -6, -5, -2$).

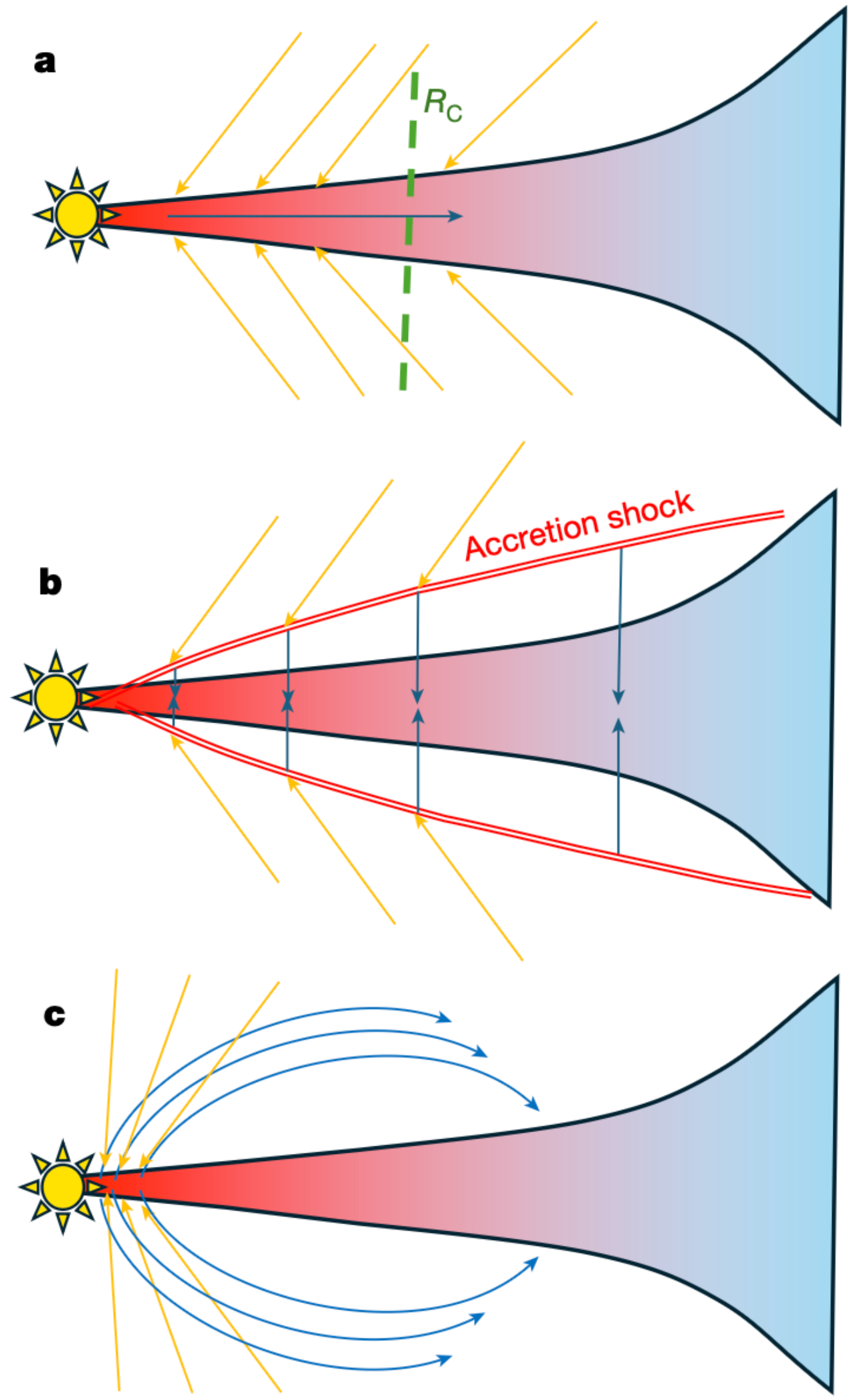

Figure 4: Possible accretion pathways of molecular-cloud gas and redistribution. Three schematic scenarios for the inflow of ISM gas into the forming solar nebula. Orange arrows indicate heated gas; blue arrows indicate cooling gas. The red–blue shapes represents the disk, with red denoting hot, high-pressure regions (>2000 K, > $10^{-3}$ bar) and blue denoting cold, low-pressure regions (<1000 K, < $10^{-4}$ bar). (a) Gas accretes inside the centrifugal radius ($R_c$), condenses near the star

at high temperature and pressure, and is redistributed by viscous spreading. (b) Infalling gas passes through a hot accretion shock, cools, and condenses before entering the disk. (c) Gas is injected hot near the star and is ejected outward, with condensation occurring during cooling in the outflow.

# Method

KineCond is a code developed to calculate the time-dependent condensation processes of minerals in the Solar Nebula, which is dominated by hydrogen. It is described in details in Supplementary Section 3 with several validation tests, and here we give a summarized version. The elements considered in the system are H, He,O, Mg, Si, Fe, Al, Na, K, Ni, Ca, Cr, S, and C. Here, the term *element* designates an atomic species. The system consists of a gas in interaction with 39 minerals. Most of the condensation codes published in the past and that allowed the investigation
of the Equilibrium Condensation Sequence [3–5,7] rely on the technics of Gibbs Free Energy Minimization (GFEM). GFEM calculates, for given P and T, the most stable combination of minerals and gas. However, like any equilibrium calculation, it provides no information if the equilibrium state is established in a reasonable time, nor does it give the list of reactions through which the equilibrium state is realized (even though some additional physical arguments may help to determine those reactions, especially at high temperature when the number of minerals in presence is small). However, each reaction has its own kinetic, that depends on P, T, and the local abundances of all elements. Thus, in order to design a time-dependent kinetic code, we must adopt a different strategy. We must explicitly specify the list of all reactions of interest, and advance each of them individually in a fully coupled way.

KineCond proceeds as follows: at each time step, we compute the number of atoms in the gas and the number of atoms in every mineral, keeping the total number of atoms constant. We assume that the pressure is constant and that only the temperature varies with time. The evolving variables of the system are: $N_i^{gas}$ (the number of moles of each molecule $i$ in the gas) and $N_j^{min}$ (the number of moles of each mineral $j$ in the system, listed in Supplementary Table 2). We assume that gas-gas reactions are much faster than gas-mineral reactions and condensation reactions, so the gas molecular composition is always close to chemical equilibrium (then the gas molecular composition only depends on the current values of T and P as well as the number of moles of each element in the gas : $N_i^{gas}$).As we focus on gas-mineral processes, reactions are divided into two broad categories: (1) condensation/evaporation reactions, and (2) gas-mineral reactions. The rates of these reactions dictate the evolution of $N_i^{gas}$ and $N_j^{min}$. The temperature varies linearly with time, dropping from 2000 K to 130 K on the time scale $T_c$ (ranging from 0.01 to 2000 years). The system evolves as follows: at each time $t$ we first calculate the molecular composition of the gas

(excluding mineral condensation) by computing the chemical equilibrium of the gas at T(t) and P and with elemental abundances $N_i^{gas}(t)$. This is performed using the iconic CEA-NASA code (which stands for ”Chemical Equilibrium with Application” distributed by NASA) [47], including about 1500 gas species in total. The instantaneous gas molecular composition is then used to compute the different condensation and gas-mineral reactions and the rate at which they proceed. The different steps of the calculation are summarized in a flow chart provided in the supplementary material (Supplementary Figure 5). We detail these calculations below.

Condensation/evaporation reactions We follow a condensation and evaporation theory developed for fortserite evaporation in a $H_2$ gas [48,49], and we generalize it to many minerals. The net formation rate of a mineral $j$ is the difference between an evaporation flux ($J_j^e$) and a condensation flux ($J_j^c$). Each of them must be computed explicitly. In the gas, the flux of any element E across a unit surface (in $moles/s/m^2$) is calculated using the kinetic theory of gases [48,49]:

$$J^c(E) = \sum_m \frac{\nu_m^E P_m}{(2\pi\mu_m RT)^{1/2}} \tag{1}$$

where $m$ is any gas molecule and $\nu_m^E$ is the stoichiometric coefficient of element E in molecule $m$, $P_m$ is the partial pressure of molecule m, $\mu_m$ the molar mass and R is the ideal gas constant. The partial pressures of gas molecules, $P_m$, are obtained by running the CEA-NASA code. We now consider a mineral $j$ with formula $\{\alpha_j^E E\}$ with $\alpha_j^E$ the stochiometric coefficient of element E in mineral $j$. The condensation flux of the mineral $j$ ($J_j^c$) is determined by the smallest flux (over all elements $E$ entering in its composition) so that :

$$J_j^c = \gamma_{E,j} min_E \left( \frac{J^c(E)}{\alpha_j^E} \right) \tag{2}$$

$\gamma_{E,j}$ is the sticking efficiency of atom E on mineral j. It ranges from 0 to 1 and is poorly constrained. Here we set $\gamma_{E,j} = \gamma = 0.1$, as a standard value, for all minerals and all atomic elements.

The evaporation flux of the mineral $j$ ($J_j^c$) is classically [50]:

$$J_j^e = \frac{P_j^{sat}}{(2\pi\mu_j RT)^{1/2}} \tag{3}$$

where $P_j^{sat}$ is the saturating vapor pressure of mineral $j$ at temperature T. $P_j^{sat}$ is easily defined in vacuum and for a mineral that has an equivalent gas form with same formula (like $H_2O$ or Fe).

However, most minerals do not have a corresponding gas form (like $Mg_2SiO_4$), making Equation 3 not directly applicable to most minerals. In addition, the presence of the surrounding gas (mainly H and He) modifies the chemical equilibrium and must be taken into account when computing the saturation vapor pressure. So, Equation 3 is not directly applicable to compute the evaporation flux of all minerals immersed in a surrounding gas rich in H. To compute $J_j^e$ from the kinetic theory of gases, we follow a strategy by which the evaporating flux is directly computed without the need to compute the saturating vapor pressure[48] and in [49]. When a mineral is in equilibrium with its surrounding gas, the evaporating and condensation fluxes exactly balance (Equations 2 and 3 ).

So, determining the evaporation flux of the mineral $j$ is equivalent to determining the condensation flux of the mineral $j$ in equilibrium with the surrounding gas at pressure P and temperature T. Following previous work[48] focusing on forsterite equilibrium with $H_2$ we have calculated the vapor composition at equilibrium with a mineral immersed in $H_2$ gas and derived the evaporative flux using equation 2. This calculation was performed using the CEA code for 39 minerals (see Supplementary Table 2). The pressure P of the $H_2$ gas varied in the range $10^{-10}$ bar $<P<$ 0.1 bar and the temperature T was varied in the range $130K < T < 2200K$ and $fO_2$ was varied by ten orders of magnitude. All of these data are compiled into lookup tables and are interpolated at current (T, P, $fO_2$). So in KineCond the ideal evaporative flux ($F_j$) from the mineral $j$ is obtained by reading a precalculated lookup table and interpolated between tabulated values at current values of $\ln(P)$, T and $\ln(fO_2)$. Examples of computed evaporative fluxes are shown in Supplementary Figure 10. The result is multiplied by $\gamma = 0.1$ so the effective evaporative flux is :

$$J_j^e = \gamma F_j(P, T, fO_2) \tag{4}$$

The time derivative of number of moles of any mineral $j$ due to competing condensation and evaporation processes is then :

$$\frac{dN_j^{min}}{dt} = S(J_j^c - J_j^e) \tag{5}$$

Where $S$ is the surface of contact of the grain with the gas, $J_j^c$ and $J_j^e$ are given by equations 2 and 4, respectively. Our code implements a first-order time solver (Euler), and during a time step $dt$, the number of moles of every mineral evolves according to Equation 5. To conserve the total number of moles of each element, the atoms released and removed from the gas ($N_i^{gas}$) are counted according to the stoichiometry of every mineral. We adopt an operator-splitting approach in which condensation reactions during the dt time step are treated first (Section ), and gas-mineral reactions are treated in a second step (Section ).

The surface of contact of minerals with the gas, S, depends on the radius of the grain ($r$) and the number of grains (N) so that $S \sim N 4\pi r^2$. The self-consistent calculation of N and r necessitates the computation of the time-dependent nucleation process and requires the taking into account of both the sticking and fragmentation processes of minerals during their settling and growth in the turbulent solar nebula. Currently, such a coupling (mineral condensation, coagulation, and fragmentation) has never been done, and some models take some processes into account (coagulation or fragmentation [51,52], or metal condensation [22]). Models coupling dust settling, coagulation, and fragmentation show that dust size distribution reaches rapidly steady-state inward 10 Astronomi-cal Unit in a few orbital periods [51,53] and that what determines dust size is rather the equilibrium between coagulation and fragmentation in turbulence, rather than mineral growth. Following these lines, and to make the calculation tractable, we have used a characteristic dust size, which gives a gas-mineral surface of contact, rather than computing self-consistently a mineral growth model, and we limit ourselves to an order-of-magnitude calculation. In other terms, we assume that all minerals (despite their mass) have a surface in contact with the gas equivalent to a sphere with radius r = 10 $\mu m$ (r is a free parameter of the model). This size is typical of the minerals observed in chondrites. The number of grains in the system, N, should be controlled by the number of nucleation sites that first appear in the gas. As aluminium is the most refractory atom in our model, we approximated $N \sim M_{Al}/(4/3\pi\rho_{Al}r^3)$ where $M_{Al}$ is the total mass of Al in our system and $\rho_{Al}$ is the density of aluminium, so S=$3M_{Al}/(\rho_{Al}r)$. This does not mean that all minerals are 10 micrometers in radius, but rather that on average the total surface of contact of a mineral is equivalent to a population of minerals with an average surface of a 10 $\mu m$ radius sphere (it could be fractal in shape). Of course, our calculation may not be accurate at the beginning of coagulation when the particle size is close to the monomer size, but changing the monomer size by a factor of 1000 only changes by 4% the growth time and does not change the final size[22]. 70-micrometer-radius particles are formed in about 10 weeks[8] (with a cooling rate of about 100 K/year at $10^{-4}$ bar, a pressure typical of the region of 0.1-1 AU), which is comparable to the orbital period at the distance of Mercury. Other works find the formation of 10 microns grains in a few orbital periods at 1 AU [51,53] or at 5 AU [52]. Of course, larger minerals can be formed, but 10 microns is typical of what is observed in meteorites.

Gas-Mineral surface reactions (nebular reactions) In addition to condensation and evaporation reactions, condensed minerals can also interact with the gas, leading to mineral transformation. We call ”nebular reactions” those reactions by which a gas interacts with a pre-existing mineral (M1) and forms a new mineral (M2) and the expanse of M1. The number of such reactions is potentially infinite and for now KineCond implements reactions with the following generic form:

$$M1 + \beta_1 E1 + \beta_3 E2 \Rightarrow \alpha M2 \quad (6)$$

Where M1 and M2 represent two minerals, E1 and E2 represent any element in gas form, and $\alpha$, $\beta_1$, $\beta_2$ are stoichiometric coefficients (normalized so that the stoichiometric coefficient of M1 is 1). For example: $Mg_2SiO_4$(s) + 1 Si + 2 O ⇒ 2 MgSiO3(s).

Elements E1 and E2 can be implied in the reaction in any molecular form (Si could be in the molecular form SiO, Si, $SiO_2$, etc.), but the radical of the molecule that is not used in the reaction is released into the gas and does not enter into the calculation of the reaction constant (due to the difference of the formation $\Delta G$ between the left and right sides of the equation, see Supplementary Sections 3.4 and 3.4.1 for a detailed calculation). For now, we are limited to reactions in which mineral M1 does not lose atoms to the gas. If M1 did, this could be an incongruent evaporation process, and this nebular reaction would be inconsistent with our hypothesis of treating condensation/evaporation as congruent processes using the Hertz-Knudsen formalism described above. Using combinatorial analysis, we found that 55 reactions of this type are possible with our mineral selections (Supplementary Table 2). After many tests, only 38 reactions were kept, the others playing a more minor or no role (Supplementary Table 3).

For a given T and for a given gas composition, we first compute if any reaction listed in Supplementary Table 3 is kinetically possible, that is, if the mineral M2 is formed (see Supplementary Section 3.4 for a detailed calculation). If the reaction favors the formation of the mineral M2, then we calculate the rate at which M2 is formed and M1 disappears.

Rate of nebular reactions For computing the reaction rate of reaction 6, we follow an approach called SCT[23] (Simple Collisional Model), but modified to take into account improvements in our understanding of reaction rates on the surface of the mineral [24]. We first determine the flux of the E1 and E2 incoming elements by summing the fluxes of all molecules in the gas, which carry elements E1 or E2 (Equations 1) called $J^c(E1)$ and $J^c(E2)$. The $c$ means *condensation*. We call them Elementary Fluxes. The smallest of the two fluxes (weighted by $\beta_1$ or $\beta_2$) controls the rate of progression of the reactions. So, if the reaction 6 is thermodynamically possible, then the rate at which mineral M1 appears (and mineral M2 disappears) is:

$$\frac{dN_{M1}^{min}}{dt} = -S \times min\left[J^c(E1)/\beta_1, J^c(E2)/\beta_2\right] \tag{7}$$

$$\frac{dN_{M2}^{min}}{dt} = -\alpha \frac{dN_{M1}^{min}}{dt} \tag{8}$$

Equation 7 assumes that all collisions lead the to reaction and overestimates the reaction rate. So, Equation 7 must be corrected. Laboratory experiments show that two regimes of reaction rates exist [24,33,54,55]: the linear regime and the parabolic regime. In the linear regime, each molecular collision has a certain probability of producing a chemical reaction, parameterized by the activation energy $E_a^l$. So the linear rate is :

$$\left.\frac{dN_{M1}^{min}}{dt}\right|_{linear} = -S \times min\left[J^c(E1)/\beta_1, J^c(E2)/\beta_2\right] e^{-E_a^l/RT} \tag{9}$$

We recover here the SCT [23]. However, laboratory experiments show that after a first period during which the reaction rim grows linearly with time at the mineral's surface [33,54], the reaction switches to a parabolic regime where atomic diffusion across the reactive rim limits the reaction rate. In this regime, the rim grows with the square root of time. In that case, the reactive layer with thickness $H$ grows like $H^2 = k(T)t$ where k(T) is a diffusion coefficient that depends on temperature ($m^2/s$) and t is time [33]. It is usual to write $k(T) = Ce^{-E_{ap}/RT}$ with $C$ and $E_a^p$ standing for a prefactor and an activation energy (in the parabolic regime). This process can be described by a Fick diffusion law where the reaction rate (in $moles/m^2/s$) is

$$\frac{dN}{dt} = -Sk(T)\frac{dc(x)}{dx} \tag{10}$$

where c(x) is the concentration of mineral M2 at location x (x = 0 corresponds to the surface of mineral M1). In order to simplify the calculation, we assume that c(x) drops linearly within M1, so that $dc/dx \sim 1/H$. So we solve :

$$\frac{dN}{dt} = -Sk(T)\frac{1}{H} \tag{11}$$

where $H$ is the thickness of the mineral M2 layer above the M1 surface and S the grain surface. In KineCond H is calculated as $H = \mu_{M2}N_{M2}/(S\rho_{M2})$ where $\mu_{M2}$ and $\rho_{M2}$ are the molar mass and density of mineral 2. To avoid nonphysical high reaction rates when H is close to 0 we bound the parabolic rate to be always smaller than the rate of incoming atoms to the M1 mineral's surface.

So, the reaction rate in the parabolic regime reads:

$$\left.\frac{dN_{M1}^{min}}{dt}\right|_{parabolic} = -S \times min\left[\frac{k(T)}{H}; min\left[J^c(E1)/\beta_1, J^c(E2)/\beta_2\right]e^{-E_a^l/RT}\right] \tag{12}$$

Unfortunately, there are only a handful of laboratory measurements [24] and most gas-mineral reactions are undocumented. For magnetite formation, laboratory experiments give $E_a^p \sim 90KJ/mol$

[54] while for Troilite (FeS) the activation energy in the parabolic rate is reported as $30KJ/mol < E_a^p < 70KJ/mol$ and in the linear regime is $28KJ/mol < E_a^l < 94KJ/mol$ [33]. These measurements were done in the (T, P) range of stability of both minerals. In contrast, the rate of forsterite to Enstatite ($Mg_2SiO_4$ + 1 Si + 2 O → $2MgSiO_3$) was measured at a temperature well above the stability of Enstatite or Forsterite (> 1700K) and $E_a{}^p \sim 500KJ/mol$ was found [56]. For magnetite, troilite, and enstatite activation energies, we use the laboratory values for the parabolic regime. For the linear regime and for all other reactions due to many uncertainties and lack of data [24], we investigate different end-member scenarios defined below:

- Fast nebular reaction : (end member)$E_a^l = 0$ and$E_p^l = 0KJ/mol$
- Moderate nebular reaction :$E_a^l = 20Kj/ml$and$E_p^l = 20KJ/mol$
- Slow nebular reactions :$E_a^l = 80Kj/mol$ and $E_a{}^p = 500KJ/mol$

The prefactor coefficient in k(T) , $C$, is determined so that the linear and parabolic fluxes connect smoothly for a rim thickness of H=1 $\mu$m. This also prevents unrealistically high reaction rates for small values of H (a well-known problem of Fick's law). So C is equal to the linear rate divided by 1 $\mu$m, that is, in the range of transition rim thicknesses measured for troilite formation [33]. For the enstatite formation experiment, the transition thickness is < $10$ microns [56].

Of course, the above procedure does not reflect the vast richness of gas-grain reactions and suffers from an important lack of experimental data. However, we have performed numerous tests, varying the values of $E_a^l$ and $E_a{}^p$. We found that changing these values does not significantly impact our results as the processes investigated here are dominated by condensation processes, rather than gas-mineral surface interactions.

Putting all things together The user first specifies the gas composition (solar in general), pressure P and cooling time $T_c$ in years. Temperature will decrease linearly from 2000K to 130K in time $T_c$. The equivalent cooling rate is therefore $(2000 - 130)/T_c$ K / year. Each time step is decomposed as follows.

1. Compute the gas molecular composition at T and P (using only atoms present in gas form) using the CEA-NASA equilibrium code [47].
2. Compute the resulting gas atomic fluxes (Equation 1).
3. Condensation/Evaporation phase: Calculate the condensation and evaporation fluxes for every mineral (Equation 5), and evolve the number of minerals accordingly and the number of elements in the gas.

4. Nebular reactions phase: Determine which nebular reactions occur in the gas (Supplementary Table 3) and compute the rate of production of mineral M2 and the rate of destruction

   of mineral M1 (Equations 9 and 12).

5. Update the number of elements in the gas, enforcing mass conservation.

6. Increment time and go back to the first step.

Several comparisons of KineCond outputs against various laboratory experiments and a comparison with the classical equilibrium condensation sequences obtained by GFEM are shown at the end of Supplementary Section 3. For all these tests, we find good, to very good, agreement.

Acknowledgments

We thank the anonymous reviewers for their comments that improved the quality of the paper. Parts of this work were supported by the DISKBUILD project (ANR-20- CE49-0006), the LabEx UnivEarthS initiative (ANR-10-LABX-0023 and ANR-18-IDEX-0001) and by the French space agency CNES (Centre National d'Etudes Spatiales). The numerical calculations were performed´ in part on the S-CAPAD/DANTE platform at IPGP. MC acknowledges funding from the ERC under the Horizon Europe program/ERC grant agreement No. 101200693 (DUST). This work was partially supported by ANR PERSEID (ANR-25-CE49-3880, PI: Yves Marrocchi). PAS was supported by the Swiss

National Science Foundation (SNSF) through an Eccellenza Professorship (203668) and the Swiss State Secretariat for Education, Research and Innovation (SERI) under contract No. MB22.00033, a SERI-funded ERC Starting grant “2ATMO”. SC thanks G. Avice, J.

Siebert and F. Moynier for fruitful discussions.

Authors contributions

S. Charnoz has designed and led the project; conceived, built and tested KineCond. All authors participated equally in the writing. J. Aleon provided his expertise on CAIs, mineral condensa-´ tion processes and helped design and testing KineCond. M. Chaussidon provided his expertise on CAIs and chondrites physics; Y. Marrocchi provided his expertise on chondrites, AOA, and chondrules. P. Sossi provided his expertise and chondrites and thermodynamics and performed the $fO_2$ equilibrium calculations using the factsage software. P. Franco performed equilibrium calculations with FASTCHEM-COND.

Additional information

High resolution mosaic images of kinetic condensation sequences displayed in poster format (A0) are available to download from the IPGP Research Collection repository with DOI number : *doi.org/10.18715/IPGP.2026.mkv2scjh*. The link to the repository is https://doi.org/10.

18715/IPGP.2026.mkv2scjh .

Data availability

The data used to produce Figures 1 to 3 were generated by the KineCond code, available in the public repository : IPGP research Collection https://doi.org/10.18715/IPGP.2026. mkv2scjh.

Code availability

A version of the KineCond code, configurated to reproduce the main results of this paper is available in the public repository : IPGP research Collection https://doi.org/10.18715/IPGP.2026.mkv2scjh.